\documentstyle[amssymb,aps,twocolumn]{revtex}

\begin{document}
\author{J. Bobroff$^{1}$, H. Alloul$^{1}$, W.A. MacFarlane$^{1}$, P.\ Mendels$^{1}$,
N.\ Blanchard$^{1}$, G.\ Collin$^{2}$, J.-F.\ Marucco$^{3}$}
\title{Persistence of Li Induced Kondo Moments in the Superconducting State of
Cuprates}
\address{$^{1}$Laboratoire de Physique des Solides, UMR 8502, Universit\'{e}\\
Paris-Sud, 91405 Orsay, France\\
$^{2}$LLB, CE-Saclay, CEA-CNRS, 91191 Gif sur Yvette, France\\
$^{3}$LEMHE, UMR 8647, Universit\'{e} Paris-Sud, 91405 Orsay, France}
\date{to appear in Phys.Rev.Lett.\ 30 april Vol.\ 86 (2001)}


\twocolumn[\hsize\textwidth\columnwidth\hsize\csname@twocolumnfalse\endcsname
\maketitle

\begin{abstract}

Using $^{7}$Li NMR shift data, the anomalous local moment induced by spinless Li impurities persists below T$_{c}$ in YBa$_{2}$Cu$_{3}$O$_{6+y}$. In the underdoped regime, the moments retain their Curie law below T$_c$. In contrast, near optimal doping, the large Kondo screening observed above T$_{c}$ (T$_{K}=135$ K) is strongly reduced below T$_{c}$ as expected theoretically when the superconducting gap develops. The limited spatial extent of the induced moment (on 1$^{st}$ near neighbour Cu) is not drastically modified below T$_{c}$, which allows a comparison with STM determination of the local density of states. Our results constrain theoretical models of the impurity electronic properties.
\end{abstract}
]

The influence of impurities on superconductors has always been used as an
effective probe of their actual properties. For example, while magnetic
impurities are the most prominent s-wave pair breakers, any type of
scattering is detrimental to d-wave superconductivity.\ To our knowledge,
experimental investigation of the modifications of the magnetic properties
of an impurity below the superconducting transition $T_{c}$ has never been
performed. Macroscopic {\em bulk} magnetic experiments are unadapted since
the various contributions to the susceptibility cannot easily be singled out
below $T_{c}$. {\em Local} measurements of the susceptibility of these
moments using hyperfine techniques are in principle possible, but are
usually prohibited by technical limitations, such as strong spin lattice
relaxation effects for the impurity NMR.

Despite this experimental void, an extensive theoretical work has been
devoted to this question in classical Fermi liquids \cite{Withoff}\cite
{Ingersent}. The behavior of the moment below $T_{c}$ is predicted to depend
both on the shape of the superconducting gap and on the Kondo temperature $%
T_{K}$ of the moment in the normal state. $T_{K}$ is a signature of the
screening of the moment by the conduction electrons and is related to its
coupling $J$ to the carriers. The primary effect, anticipated, but not
observed so far, is a reduction of the Kondo screening due to the pairing of
the carriers. For small $J$, this results in a complete restoration at low $%
T $ of the Curie susceptibility of the moment.

In cuprate superconductors, which are correlated electronic systems, the
magnetic properties of impurities are more intricate. In the normal state,
spinless impurities like Zn \cite{mahajanPRL94}\cite{Mendels}\ or Li \cite
{BobroffPRLLi}\cite{AMF} substituted on the copper site of the CuO$_{2}$
planes induce a local moment in their vicinity. This moment extends
essentially on the four Cu near neighbours (n.n.) of the impurity. Its
static \cite{BobroffPRLLi} and dynamic \cite{AMF} susceptibilities exhibit a
Kondo like behaviour with a large range of $T_{K}$ values, which can be
spanned by changing hole doping.\ {\em The effect of superconductivity on
this moment addresses two issues:} {\em the persistence of magnetic
correlations in the superconducting state, and the influence of d-wave
pairing on the Kondo screening}.

We present here the first measurements of the induced moment properties
below $T_{c}$.\ they are performed using $^{7}$Li NMR since the transferred
hyperfine couplings are weak enough that relaxation effects do not prohibit
NMR spectroscopy of the moment. We propose a method to extract the
susceptibility $\chi _{loc}$ probed at Li sites.\ This requires correcting
the internal field seen by Li for screening and vortex effects due to
superconductivity.\ It will then be shown that the Li induced moments
survive below $T_{c}$ and that $T_{K}$ is strongly reduced. Furthermore, we
will demonstrate that these induced moments remain confined primarily to the 
$1^{st}$ n.n. coppers below $T_{c}$.\ Recent scanning tunneling microscopy
(STM)\ experiments in the Zn substituted Bi2212 cuprate gave a measure of
the local density of states (LDOS) in the superconducting state \cite
{PanNature}.\ The STM data suggest the occurence of a LDOS peak near the
Fermi level on the Zn site and on the 2$^{nd}$ n.n. Cu.\ This location
contrasts with our finding of a magnetic state located dominantly on the $%
1^{st}$ n.n. Cu. The discussion of this discrepancy will lead us to favor
theoretical models which incorporate the magnetic character of the impurity.

The Li substituted samples YBa$_{2}$(Cu$_{1-x_{n}}$Li$_{x_{n}}$)$_{3}$O$%
_{6+y}$ are those used in \cite{BobroffPRLLi}.\ The two batches with Li
nominal concentrations $x_{n}=1\%$ and $2\%$ had an effective in-plane Li
concentration of $0.85\%$ and $1.86\%$ per CuO$_{2}$ layer. Two oxygen
contents were obtained from each batch corresponding to optimally ($y=0.97$)
and under- ($y=0.6$) doped regimes. Their $T_{c}$ were found to be $85.3$ K
and $79.5$ K at optimal doping, and $41$ K and $25$ K for the underdoped
materials.\ The sample crystallites were aligned along the $c$
crystallographic axis with an applied field.\ This allows accurate NMR
measurements performed in fields parallel to $c$ ranging from 3 to 7 Tesla.
Below $T_{c}$, NMR\ spectra are too broad for Fourier Transform spectroscopy
and were then measured point per point by sweeping the frequency over a few
hundreds of kHz \cite{NoteSweepField}.

In the superconducting state, the internal field $B$ at any point in the
sample may differ from the applied field $B_{app}$ due to large screening
effects.\ NMR measurements of the spin contribution to the NMR\ shift, which
is proportional to $\chi _{loc}$ are not straightforward. The spectral
position $\omega ^{\ast }$of the NMR\ signal of a $^{7}$Li nucleus is
related to its NMR shift $^{7}K$ by

\begin{equation}
\omega ^{\ast }=\gamma (B_{app}+\delta B)(1+^{7}K)  \label{formuleShift}
\end{equation}
with $\delta B=B-B_{app}$, and $\gamma $ the nuclear gyromagnetic factor. An
independent determination of the distribution of $\delta B$ at Li sites is
needed.\ We can use the fact that the average of $\delta B$ is almost
independent of $B_{app}$ in a large range of fields.\ Indeed, the
magnetization $M\varpropto \delta B$ has been measured to be flat from $%
B_{app}=2$\ to $12$ Tesla in similar pure powder ceramics \cite{senoussi}.
Measurements for two values of $B_{app}$ yield both $\delta B$ and $K$ at
the Li sites from Equ.\ref{formuleShift} applied to the peak position of the
NMR.

Let us specify first how $\omega ^{\ast }$ has been extracted from the NMR\
spectra such as those plotted in Fig.1. They consist of three lines at $%
\omega ^{\ast }-\omega _{c}$, $\omega ^{\ast }$, and $\omega ^{\ast }+\omega
_{c}$, due to the quadrupolar splitting of the $I=3/2$ Zeeman transitions by
the electric field gradient (EFG) at the Li site. The quadrupolar frequency $%
\omega _{c}$ is proportional to the EFG in the direction $c$ of $B_{app}$ 
\cite{BobroffPRLLi}.\ Above $T_{c}$, the two outer lines are broader than
the central one due to a small distribution of $\omega _{c}$ (typically $%
\delta \omega _{c}/\omega _{c}\sim 10\%$).\ The other source of broadening,
common to the three lines is the local distribution of hole content which
induces a slight distribution of $^{7}K$.\ It scales with $^{7}K$\ and
therefore increases with decreasing $T$.\ \ Below $T_{c}$ the presence of
pinned vortices induces a distribution of $\delta B$ amoung Li nuclei,
leading to an additionnal asymmetric broadening (Eq.\ref{formuleShift}), as
observed in Fig.1. The high frequency tail and the center peak of the line
correspond respectively to Li sites in the vortex cores with $\delta B>0$,
and between vortices with $\delta B<0$ \cite{Brandt}. We fitted the $T<T_{c}$
spectra using the $T>T_{c}$ shape convoluted by an asymmetric gaussian
representing the $\delta B$ distribution.\ The resulting values of $\delta B$
using two measurements of the central line peak position $\omega ^{\ast }$
are plotted in Fig.2 for optimal doping. The negative sign of $\delta B$
confirms that {\em the susceptibility probed by the peak position }$\omega
^{\ast }${\em \ is associated to Li defects between vortices in the bulk
superconducting state. }At low $T$, the obtained $\delta B\simeq -30$ G is
consistent with measurements by $\mu $SR\ or $^{89}$Y\ NMR in pure compounds 
\cite{BarretRiseman}. The $T$-dependence of $\delta B$ originates from the $%
T $ variation of the superconducting screening and of the field distribution
in the vortex network. The data for $\delta B(T)$ has been fitted by a
phenomenological power law corresponding respectively for $x_{n}=1\%$ and $%
2\%$ to $\delta B_{1\%}=-32(1-(T/77)^{2})$ and $\delta
B_{2\%}=-26(1-(T/60)^{1.5})$ in Gauss units. The decrease of $\delta B$ with
Li content can be explained by the concomitent increase of the penetration
depth $\lambda $ measured by $\mu SR$ \cite{MusrMendels}. Above 77\ and 60K\
respectively, we found $\delta B=0$ within error bars ($\pm $4 G). For
optimally doped samples and applied fields of a few Tesla, it has been shown
that the vortices are in a liquid state in our range of temperatures below $%
T_{c}$ \cite{Reyes}\cite{Sonier}.\ Each vortex is then moving much faster
than the NMR timescale, which averages out both the broadening and the
screening effects, leading to $\delta B=0$. The exact determination of the
melting temperature, which depends on $B_{app}$, $x_{n}$ as well as on the
sample microstructure, is beyond the scope of this work. In the underdoped
compound, for $B_{app}=7$ T and $T>10\;$K, the vortices should always be in
the liquid state \cite{Sonier}\cite{Andresson}.\ Indeed we find that $\delta
B=0$ within experimental accuracy, so that no correction to $\omega ^{\ast }$
was needed for such conditions.

Using this determination of $\delta B$ in Eq.\ref{formuleShift}, the shift $%
^{7}K$ can then be safely extracted.\ As seen in Fig.1, this shift and
therefore $\chi _{loc}$ are increasing with decreasing $T$ below $T_{c}$.
This is confirmed by systematic measurements of $^{7}K$ for all
concentrations of Li and oxygen dopings represented in Fig.3.\ A more
compelling representation of the variations of $^{7}K$ is obtained by
plotting $1/(^{7}K-^{7}K_{0})\sim 1/\chi _{loc}$ versus $T$ as done in
Fig.4. $^{7}K_{0}$ is the $T$ independent part of the shift and is measured
from high $T$ data to be much smaller than the observed variations of $^{7}K$%
. In this plot the Curie-Weiss law $^{7}K-^{7}K_{0}=C/(T+T_{K}),$ which
represents $\chi _{loc}$ in the normal state is a straight line with slope $%
C^{-1}$ which intercepts the horizontal axis at $-T_{K}$.

In the underdoped regime, from both Fig.3 and 4, no significant change
occurs at $T_{c}$, i.e the almost perfect Curie law observed above $T=80$ K
is not affected by superconductivity and corresponds to $T_{K}=2.8\pm 1$ K 
\cite{NoteMesuresLowT}. To our knowledge, this is {\em the first measurement
of }$\chi _{loc}${\em \ in a superconductor for moments appreciably coupled
to the carriers}. In contrast we find a sharp increase of $\chi _{loc}$
below $T_{c}$ at optimal doping, as seen in the lower panel of Fig.3.\ This
increase is not expected for a usual Kondo impurity. This can be seen by
comparing our data with a prototype of the Kondo susceptibility such as $%
\chi _{Fe}$ of Fe impurities in the dilute alloy {\bf Cu}Fe. We have scaled
the $T$ variation of $\chi _{Fe}$ measured in \cite{Alloul} by a factor $%
T_{K}($O$_{7})/T_{K}(${\bf Cu}Fe$)\simeq 135/27.6\simeq 4.9$\ to fit the
normal state data. One can obviously see in Fig.3 and 4$\;$that the rescaled 
$\chi _{Fe}$ saturates at low T like $\chi _{loc}$ but at a much lower value
(of about a factor three). The data for $\chi _{loc}(T<T_{c})$ is better
fitted by another scaling of $\chi _{Fe}$ also represented in Fig.3 and 4,
leading to $T_{K}=41\pm 7$ K instead of $135$ K above T$_{c}$. {\em The
moments survive below }$T_{c}$ {\em even at optimal doping.\ They still
display a Kondo-like susceptibility with a weaker screening than in the
normal state}. This value of $T_{K}$ is consistent with the analysis of the
specific heat measurements of Zn substituted YBaCuO$_{7}$ by Sisson {\it et
al.}\cite{sisson} who attributed the absence of a Shottky anomaly below $%
T_{c}$ to Kondo screening of the Zn induced moments.

In the superconducting state of a Fermi Liquid, the decrease of the DOS\ at
the Fermi level prohibits the development of the Kondo divergence near or
below a critical coupling $J_{c}$.\ This is also true for d-wave
superconductors for which the gap corresponds to a linear energy dependence
of the DOS \cite{Withoff}\cite{Ingersent}, except at low T in the presence
of impurities. Qualitatively this explains the reduction of Kondo screening
and $T_{K}$ at optimal doping below $T_{c}$.\ Renormalization group
numerical studies using a realistic set of parameters may quantitatively
account for the behavior observed in Fig.\ 4 at optimal doping \cite
{Ingersent}.

In the underdoped regime, the absence of detectable modification of $\chi
_{loc}$ below $T_{c}$ could result from the already small value of $T_{K}$
found in the normal state. In contrast with optimal doping, any reduction of 
$T_{K}$ below $T_{c}$ cannot be observed in our experimental conditions
where $T\gg T_{K}$. The low value of $T_{K}$ already in the normal state
could be explained by the occurence of the pseudogap, similar to the effect
of the superconducting gap at optimal doping. Both gaps display the same
d-wave symmetry \cite{photoemission}, and should lead to a reduction of $%
T_{K}$.\ However, the significant difference between the optimal and
underdoped regimes in the apparent $T_{K}\,\ $(respectively $41$ K and $2.8$
K) in the superconducting regime remain to be understood.

Even though the above Fermi liquid picture explains the $T$ behavior of $%
\chi _{loc}$, it cannot account for the very existence of the moment induced
by spinless impurities.\ This moment is the result of electronic
correlations intrinsic to the pure cuprate.\ From NMR experiments performed
in the normal state, this moment consists of a staggered AF state which
extends on many lattice sites, but resides predominantly on the impurity 1$%
^{st}$\ n.n Cu sites. The present experiment demonstrates that this is still
true below $T_{c}$. At optimal doping, scaling with $\chi _{Fe}$ yields
numerical values for the Curie term $C$ of $5.1$ $10^{4}$ K.ppm and $4.8\pm
1.1$ $10^{4}$ K.ppm above and below $T_{c}$ respectively.\ These values are
only 25\%\ smaller than in the underdoped case.\ Hence $C$ is almost
unaffected either by superconductivity or hole doping. {\em Therefore the
interaction with the charge carriers does not modify the effective moment on
the 1}$^{st}${\em \ n.n. but merely induces a modification of Kondo screening%
}.

The measurements by STM on Zn substituted Bi2212 \cite{PanNature} also imply
a small spatial extent of the impurity LDOS below $T_{c}$. This LDOS
exhibits a narrow resonance peak at an energy of 1.5 meV ($\approx $18K)
below the Fermi level.\ This energy scale is close to our $T_{K}$ value,
suggesting a common origin for the two phenomena. But the absence of LDOS on
the 1$^{st}$ n.n. found by STM contrasts with our finding. Most computations
of the LDOS can reproduce the symmetry of the observed STM\ pattern, but did
not consider the existence of an induced magnetism \cite{theoriesSTM}\cite
{Haas}.\ In order to explain altogether the magnetism and the STM
experiments, we suggest two possible scenarii:

\ i) For an on site potential, some interpretations \cite{theoriesSTM}
predict a weak LDOS on the impurity site, and an enhanced LDOS on the 1$^{st}
$ n.n., which might agree with NMR.\ If, as usually assumed, the tunneling
is vertical from the Bi to the Zn site, this\ prediction would result in an
observation of a large LDOS on the 1$^{st}$ n.n., in contrast with STM data.
\ We suggest that the tunneling occuring through a BiO layer could have a
much larger matrix element to the n.n. sites than to the one on the
vertical.\ This would yield an {\em apparent} LDOS on-site and on the 2$^{nd}
$ n.n in the STM experiments \cite{finale}.

ii) Let us assume that the perturbation induced by the impurity corresponds
to a potential on its n.n.\ Cu.\ We can anticipate that calculations along
the lines of \cite{theoriesSTM}, with a potential on a n.n. Cu, would induce
an LDOS on the impurity and the 2$^{nd}$ n.n. Cu sites.\ The total {\em %
actual} LDOS would then be consistent with STM\ for a vertical tunneling.
Polkovnikov {\it et al.}\ have done such a computation \cite{Sachdev}, which
is furthermore the first realistic attempt to account for the magnetic
properties detected by NMR. They introduced an extended moment on the four\
Cu n.n. sites, exchange coupled to the quasiparticles of the superconductor
and could reproduce the spatial dependence of the LDOS observed by STM.\ 

In conclusion our measurements show that the moment induced by spinless Li
in the normal state of cuprates still exists in the superconducting state.
When the Kondo temperature $T_{K}$ is comparable to$\,T_{c}$, the Kondo
screening is strongly reduced below $T_{c}$.\ This is consistent with
computations made in a classical Fermi liquid in the presence of a d-wave
gap.\ This reinforces the analogy of the magnetic behaviour with that of a
classical Kondo effect.

We have found that the dominant magnetic contribution still resides on the 1$%
^{st}$ n.n. Cu of the impurity below $T_{c}$.\ Therefore the short range AF
correlations remain in the superconducting state. The present work leads us
to believe in a common understanding of the local magnetism and LDOS, in the
spirit of \cite{Sachdev}. Determination of the evolution of the LDOS with
temperature and hole doping should help to establish the relationship
between NMR and STM results.\ This would also constrain the theoretical
models, which should in addition account for the low T magnetic
susceptibility using the energy dependence of the LDOS.

We acknowledge N.\ Bulut, B.\ Coqblin, L.\ Fruchter, M.\ Flatt\'{e}, M.\
Gabay, A.\ Georges, S.\ Haas, K.\ Ingersent, S.\ Sachdev, S.Senoussi, M.\
Vojta for fruitful discussions.

\begin{figure}[tbp]
\caption[1]{$^{7}$Li NMR spectra for YBa$_{2}$Cu$_{3}$O$_{6.97}$ with $%
x_{n}=1\%$ obtained either by Fourier Transform or point by point. In the
superconducting state $^{7}K$ has been obtained after correction of
demagnetization effects using Eq.\ref{formuleShift}. }
\label{fig.1}
\end{figure}

\begin{figure}[tbp]
\caption[2]{Difference $\protect\delta B$ between applied and internal field
on Li sites in optimally doped compounds deduced from measurements in $%
B_{app}=3\ $and $7 $ Tesla. Straight and dot lines are phenomenological fits
given in the text. }
\label{fig.2}
\end{figure}

\begin{figure}[tbp]
\caption[3]{$T$ variation of the $^{7}$Li NMR shift $^{7}K$ . The arrows
indicate $T_{c}$ in a 30 G applied field. For the underdoped O$_{6.6}$\
sample, the full line is a Curie-Weiss fit with $C=6.5$ $10^{4}$ $ppm.$K and 
$T_{K}=2.8$ K. For the optimally doped O$_{6.97}$ samples , the prototype
Kondo susceptibility of Fe in {\bf Cu}Fe from Ref.\protect\cite{Alloul}
scaled to fit our data above (below) T$_{c}$ is represented by a full
(dashed) line. }
\label{fig.3}
\end{figure}

\begin{figure}[tbp]
\caption[4]{$T$ variation of the inverse of $^{7}K(T)-K_{0}$ which
represents the inverse of the local susceptibility $\protect\chi _{loc}$of
the induced moments nearby Li. Solid triangles correspond to underdoped $%
x_{n}=1\%$ YBa$_{2}$Cu$_{3}$O$_{6+y}$ while solid (empty) circles correspond
to $x_{n}=1\%$ ($2\%$) at optimal doping. The fits with the {\bf Cu}Fe Kondo
susceptibility of fig.3 are plotted with the same symbols. }
\label{fig.4}
\end{figure}


\begin{references}
\bibitem{NoteSweepField}  The equivalent measurement by sweeping the applied
field is not possible here since the presence of frozen vortices makes the
internal field insensitive to a change of the applied field \cite{Sonier}.
\bibitem{NoteMesuresLowT}  Measurements below T=$15$ K are not reported as
the accuracy is reduced by the large broadening of the NMR line together
with the overlap with a sharp line corresponding to Li near a chain site.
\end{references}
\end{document}